# Informetric Analyses of Knowledge Organization Systems (KOSs)

Wolfgang G. Stock

Head of the Information Science Department

Heinrich Heine University Düsseldorf

Düsseldorf, Germany

stock@phil.hhu.de

## Abstract

*A knowledge organization system (KOS) is made up of concepts and semantic relations between the concepts which represent a knowledge domain terminologically. We distinguish between five approaches to KOSs: nomenclatures, classification systems, thesauri, ontologies and, as a borderline case of KOSs, folksonomies. The research question of this paper is: How can we informetrically analyze the effectiveness of KOSs? Quantitative informetric measures and indicators allow for the description, for comparative analyses as well as for evaluation of KOSs and their quality. We describe the state of the art of KOS evaluation. Most of the evaluation studies found in the literature are about ontologies. We introduce measures of the structure of KOSs (e.g., groundedness, tangledness, fan-out factor, or granularity) and indicators of KOS quality (completeness, consistency, overlap, and use).*

## Introduction

For many evaluation models, an important indicator of the quality of an information service is the quality of its content. In our Information Service Evaluation (ISE) model (Schumann & Stock, 2014), Content Quality is a sub-dimension of the main dimension of Information Service Quality (see Figure 1). The content quality concentrates on the knowledge that is stored in the system (DeLone & McLean, 1992; DeLone & McLean, 2003; Jennex & Olfman, 2006). Knowledge in regard to information services consists of two aspects, namely the knowledge of the documents (the knowledge authors put into their publications) and knowledge of the surrogates (the knowledge indexers put into the document's metadata). In turn, the knowledge of the surrogates has two dimensions: the quality of indexing (applying the right concepts to describe the document's knowledge; Stock & Stock, 2013, p. 817-825) and the quality of the Knowledge Organization System (KOS), which is deployed for indexing (Stock & Stock, 2013, p. 809-816). A KOS is an order of concepts which is used to represent (in most cases: scientific or other specialized) documents. Common types of KOSs include nomenclatures, classification systems, thesauri and ontologies. KOSs are applied in professional information services which support scholarly communication by the provision of specialized literature. While there is a vast number of studies on indexing quality and its indicators (e.g., indexing depths including indexing exhaustivity of a surrogate and indexing specificity of the attributed concepts, indexing effectivity of the concepts, and indexing consistency of the surrogates), there are in information science only few works on the quality of the KOSs.

Information services for science and technology (e.g., *Medline* for medicine, *Chemical Abstracts Service* for chemistry or *Inspec* for physics) and information services in the context of corporate knowledge management in many cases apply so-called "controlled vocabularies" or "documentation (or documentary) languages" for the purposes of information indexing and information retrieval. Such vocabularies organize the concepts and the semantic relations betweens the concepts of a specific knowledge domain in a "Knowledge Organization System" (KOS).

The aim of this article is to underline the importance of the evaluation of KOSs as part of empirical information science, i.e. informetrics. According to Tague-Sutcliffe, informetrics is "the study of the quantitative aspects of information in any form … and in any social group" (Tague-Sutcliffe, 1992, p. 1). Wolfram divides informetrics in two aspects, namely "system-based characteristics that arise from the documentary content of IR systems and how they are indexed, and usage-based characteristics that arise from the way users interact with system content and the system interfaces that provide access to the content" (Wolfram, 2003, p. 6). Stock and Weber (2006) distinguish three subjects and accordingly three research areas of informetrics: (1) information users and information usage (with the area of user/usage research); (2) information itself including special information (e.g., science information) and Web information (with the research areas of bibliometrics, scientometrics and webometrics); (3) information systems (with the research area of evaluation and technology acceptance studies). The informetric analysis of KOSs is part of Wolfram's system-based characteristics and of Stock and Weber's information systems evaluation research.

Evaluation studies are able to answer two questions (Drucker, 1963): Do we do the right things (leading to an evaluation of effectiveness), and do we do the things in a right way (this time leading to an evaluation of efficiency)? Concerning KOS evaluation, effectiveness means the construction of right KOS, and efficiency the appropriate construction of the KOSs (adequately employed funds, speed of implementation, optimal software tools, etc.) (Casellas,

2009, p. 597). We focus on effectiveness and ignore efficiency. Our research question is: *How can we informetrically analyze the effectiveness of KOSs?* Quantitative informetric indicators allow for the empirical description, for comparative analyses as well as for evaluation of KOSs and their quality. With the empirical investigation of KOSs we break new ground in the theories of informetrics.

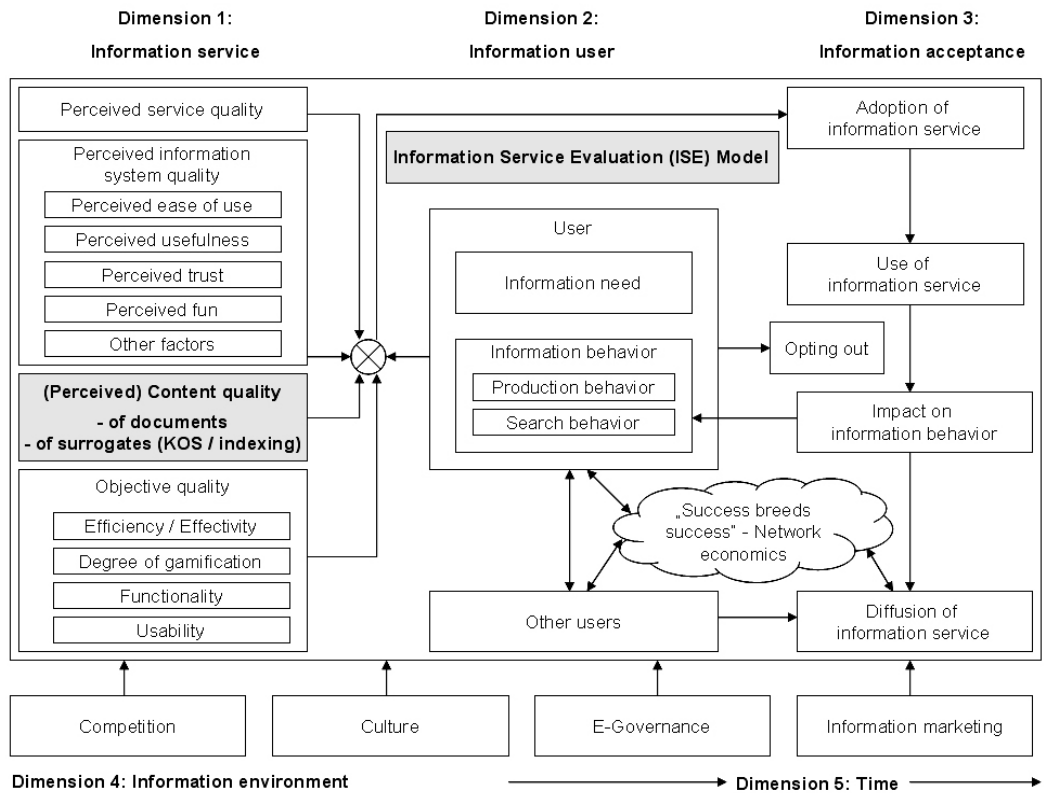

**Figure 1:** The Content Quality Dimension in the Information Service Evaluation (ISE) Model. *Source:* Schumann & Stock, 2014, p. 8 (modified).

In the next paragraph, we are going to describe very briefly Knowledge Organization Systems as systems of concepts and semantic relations. Hereafter, an overview on the state of the art of the description and evaluation of KOSs will follow. In the article's core paragraph, we present measures and indicators for the informetric evaluation of KOSs. The aim is not only to present a synthesis of a large number of approaches of KOS evaluation, but also to propose a solution for a comprehensive set of basic KOS structure measures and of KOS assessment criteria. For KOS developers, these measures and indicators should provide useful hints to construct good nomenclatures, thesauri, classification systems, and ontologies.

**Table 1:** Knowledge Organization Systems (KOSs) and the Relations They Use. *Source:* Stock, 2010, p. 1965 (modified).

|  | *Folksonomy* | *Nomen-clature* | *Classifi-cation* | *Thesaurus* | *Ontology* |
|---|---|---|---|---|---|
|  | Tag | Keyword | Notation | Descriptor | Concept |
| *Equivalence* | - | yes | yes | yes | yes |
| - *Synonymy* | - | yes | yes | yes | yes |
| - *Gen-identity* | - | yes | yes | - | yes |
| *Hierarchy* | - | - | yes | yes | yes |
| - *Hyponymy* | - | - | - | yes | yes |
| - *Meronymy* | - | - | - | yes | yes |
| - *Instance* | - | - | - | as req. | yes |
| *Further relations* | - | - | - | yes | yes |
| - *"See also"* | - | as req. | as req. | yes | yes |
| - *Specific relations* | - | - | - | - | yes |
| *Syntagmatic relation* | yes | yes | yes | yes | no |

## Concepts and Semantic Relations

Knowledge Organization Systems consist of both concepts as well as semantic relations between the concepts with respect to a knowledge domain (Stock, 2010). A "concept" is a class containing certain objects as elements where the objects have certain properties. The linguistic expression of a concept is a "word." Concepts do not exist independently of one another, but are interlinked. We will call relations between concepts "semantic relations" (Khoo & Na, 2006; Storey, 1993). Apart from folksonomies, semantic relations in KOSs always are "paradigmatic" relations, i.e., relations which are valid independently of documents (in contrast to syntagmatic relations, which depend on co-occurrences of concepts in documents). In KOSs, the following semantic relations are important:

- Equivalence (synonymy, quasi-synonymy, or gen-identity between concepts);
- Hierarchy (hyponymy, meronymy, and instance);
- and as a residual class, Further relations ("see also" as association relation, or specific relations such as usefulness or has_subsidiary_company in an enterprise KOS).

We define knowledge organization systems via their cardinality for expressing concepts and semantic relations. The three "classical" methods in information science and practice— nomenclature, classification, thesaurus—are supplemented by folksonomies and ontologies. Folksonomies represent a borderline case of KOSs, as they do not have a single paradigmatic relation (Peters, 2009).

Nomenclatures (keyword systems) distinguish themselves mainly by using the equivalence relation and ignoring all forms of hierarchical relation. In classification systems, the (unspecifically designed) hierarchy relation is added. Thesauri also work with hierarchy; some use the unspecific hierarchy relation, others differentiate via hyponymy ("is-a" relation) and meronymy ("part-of" relation). In thesauri, a generally unspecifically designed associative relation ("see also") is necessarily added. Ontologies make use of all the paradigmatic relations mentioned above. They are modeled in formal languages, where terminological logic is also accorded its due consideration. Compared to other KOSs, ontologies categorically contain instances (individual concepts). Most ontologies work with precisely defined, further relations. The fact that ontologies directly represent knowledge (and not merely the documents containing the knowledge) allows the syntagmatic relations to disappear in this case.

**State of the Art of the Evaluation of KOSs**

Most of the evaluation studies found in the literature are about ontologies (for overview articles, see Brank, Grobelnik, & Mladenić, 2005; Gangemi, Catenacci, Ciaramita, & Lehmann, 2005; Gómez-Pérez, 2004a; Hartmann et al., 2005; Kehagias, Papadimitriou, Hois, Tzovaras, & Bateman, 2008; Obrst, Ceusters, Mani, Ray, & Smith, 2007; Pak & Zhou, 2011; Vrandečić, 2009). The first article on KOS evaluation—by Gómez-Pérez in 1995— was on ontologies as well. Our scope is broader and covers all kinds of KOSs. There are a few evaluation studies about other kinds of KOSs. Vogel (2002) developed a set of quality criteria for classification systems and thesauri deployed in his retrieval system *Convera*. Applying parameters such as usability, scope, recall and precision, Owens and Cochrane (2004) worked out methods of thesaurus evaluation. Wang, Khoo and Chaudhry (2014) evaluated the navigation effectiveness of a classification system.

Gómez-Pérez, Fernández-López and Corcho (2004) distinguish between KOS verification and KOS validation. While verification is focused on the correct (formal as well as informal) representation of concepts and semantic relations (with aspects like consistency, completeness and redundancy; Lovrenčić & Čubrillo, 2008), KOS validation refers to the "real world," i.e., the comparison between the content of the KOS and its "real" counterpart in the corresponding knowledge domain (Lovrenčić & Čubrillo, 2008).

Based on the definition given by Sabou and Fernandez (2012, p. 194), KOS evaluation is the determination of the quality of a KOS against a frame of reference. In this definition, there are two crucial concepts. What is the definition of "quality" and to what does

"frame of reference" refer? *Quality criteria* define a "good" KOS. Vrandečić (2009, p. 295-296) provides us with a list of such quality criteria, among others: accuracy (Does the KOS correctly represent its knowledge domain?), adaptability (Does the KOS anticipate is use?), completeness (Is the knowledge domain appropriate covered?), consistency (Is the KOS logically coherent?) and commercial accessibility (Is the KOS easy to access and to deploy?). But all these quality criteria are "desiderata, goals to guide the creation and evaluation of the ontology. None of them can be directly measured" (Vrandečić, 2009, p. 296). It is important to have in mind that we cannot always work with *quality measures*, but only with *quality indicators*. There are several frames in the literature concerning the *frame of reference* (Brank, Grobelnik, & Mladenić, 2005; Sabou & Fernandez, 2012, p. 197 ff.), namely primitive metrics (as, e.g., the number of concepts), data-driven frames (comparisons of the KOS with its knowledge domain), KOS to KOS comparisons, frameworks concerning the syntactic and semantic structures of the KOS and, finally, user-driven frames (experiments with users and questionnaires or interviews). For all frames of reference, we present illustrative examples from the literature.

*Primitive Metrics*

Simple evaluation metrics—better known as description metrics—are based on counting concepts and relations in KOSs. For Huang and Diao (2006, p. 133), the "Concept Quantity evaluation is to count the number of concepts in the ontology," and the "Property Expectation evaluation provides an overview of the abundance of relations between concepts." Tartir and Arpinar (2007) distinguish between inheritance relationships (relations, in which the concepts' properties become inherited to the concepts' narrower terms such as the hyponymy) and other relations (as, e.g., the association relation) and count both kinds of relations. Additionally, Tartir and Arpinar (2007, p. 187) count instances, otherwise known as concepts in the KOS which represent individuals. Yang, Zhang and Ye (2006, p.165) work with the average number of relations per concept. A more subtle indicator is the "tree balance" of the KOS (Huang & Diao, 2006, p. 133): "If a tree is balanced, all its sub-trees have the same structure."

*KOSs and Their Knowledge Domains*

Does a KOS represent its knowledge domain adequately? Brewster, Alani, Dasmahapatra and Wilks (2004) re-define the well-known recall and precision metrics with regard to KOSs:

> One would like *precision* to reflect the amount of knowledge correctly identified (in the ontology) with respect to the whole knowledge available in the ontology. One would like to define *recall* to reflect the amount of knowledge correctly identified with respect to all the knowledge that it should identify. (Brewster et al., 2004, p. 1)

For the authors, "knowledge" refers to "concepts," as represented linguistically by words. They developed a corpus of typical documents for the knowledge domain and compared the words in the texts with the words in the KOS. For Brewster and colleagues the ontology "can be penalized for terms present in the corpus and absent in ontology, and for terms present in the ontology but absent in the corpus" (Brewster et al., 2004, p. 3). While it is not difficult to identify the words in a KOS, it is a bold venture to collect typical (or even all) documents of the given knowledge domain.

*KOS and Other KOSs*

To indicate the uniqueness of a KOS, it is necessary to compare it with other KOSs. The simple research question here is: "So, how may we measure the similarity of ontologies or of ontology parts?" (Maedche & Staab, 2002, p. 251). But the answer is by no means as simple as the question. In the literature, there are two approaches to study similarity between KOSs: one approach based on common words and concepts in the vocabulary (Maedche & Staab, 2002; Obrst et al., 2007, p. 146-147; Brank, Grobelnik, & Mladenić, 2005), and another that works with indexed documents in the case of polyrepresentation (i.e., applying different KOSs to index the same documents) (Haustein & Peters, 2012).

*Syntactic and Semantic Structure of KOSs*

The evaluation of the syntactical structure is targeted at the correct use of a formal language. For ontologies assigned for the application in the semantic web, the Web Ontology Language (OWL) and Resource Description Framework (RDF) are used. Much more important is the evaluation of the semantic structure of a KOS. Fahad and Abdul Qadir (2008) distinguish between redundancy, incompleteness (which is similar to the data-driven approach of recall and precision), and inconsistency. Redundancy occurs when certain information is inferred more than once in the KOS, for example, when a concept is located twice in the KOS at two different positions. Inconsistency is mainly the consequence of circularity errors (a concept is defined as a broader term or as a narrower term of itself) and partition errors (wrong decompositions of a concept into narrower terms). Fahad, Abdul Qadir and Noshairwan

(2007) determine that "the main reason for these errors is that ontologists do not classify the concepts properly" (p. 286).

*User-driven Approaches*

Noy (2004) calls indicators like completeness, consistency, and correctness "objective" evaluation criteria: "Although all these evaluation types or comparison methods are necessary, none are helpful to *ontology consumers*, who need to discover which ontologies exist and, more important, which one would be suitable for their tasks at hand" (Noy, 2004, p. 80). To get a user-driven impression of the quality of KOSs, some authors conducted experimental studies with test persons or interviewed users with the aid of questionnaires or guides.

Casellas (2009) evaluated a KOS through usability measures. He offered two questionnaires, one with questions concerning concepts, definitions, instances, and relations of the KOS, and a second one with more general items (as, e.g., "I found the ontology easy to understand," or "I thought there was too much inconsistency in this ontology."). The questionees were experts in the knowledge domain of the KOS. They were asked to express their opinions on a scale between 0 and 5 (first questionnaire) and 1 and 10 (second questionnaire).

Suomela and Kekäläinen (2006) evaluated an ontology as a query construction tool. Wang, Khoo and Chaudhry (2014) evaluated the navigational effectiveness of a classification system. Both studies worked with experiments (task-based test method) as well as with interviews (Wang, Khoo, & Chaudhry, 2014) or questionnaires (Suomela & Kekäläinen, 2006).

**Measures and Indicators of the Evaluation of KOSs**

In this section, we introduce informetric measures and indicators of KOS evaluation. Based upon the literature review and the chapter on evaluation of KOSs in our *Handbook of Information Science* (Stock & Stock, 2013), we introduce one set of measures of the structure of KOSs and four indicators of KOS quality (completeness, consistency, overlap, and use).

*Basic Structure Measures*

Several simple parameters can be used to analyze the structure of a KOS (Gangemi, Catenacci, Ciaramita, & Lehmann, 2006). These parameters relate both to the concepts and to the semantic relations. We will introduce the following structural measures:

- Number of concepts;

- Semantic expressiveness (number and kind of semantic relations);

- Granularity (average number of semantic relations per concept);

- Number of hierarchy levels;

- Fan-out factor (number of top terms);

- Groundedness (number of bottom terms);

- Tangledness (degree of polyhierarchy); and

- Precombination degree (average number of partial terms per concept).

An initial base value is the number of concepts in the KOS. Here the very opposite of the dictum "the more the better" applies. Rather, the objective is to arrive at an optimal value of the number of terms that adequately represent the knowledge domain and the documents contained therein, respectively. If there are too few terms, not all aspects of the knowledge domain can be selectively described. If a user does not even find "his/her" search term, this will have negative consequences for the recall, and if he/she does find a suitable hyponym, the precision of the search results will suffer. If too many concepts have been admitted into the KOS, there is a danger that users will lose focus and that only very few documents will be retrieved for each concept. When documents are indexed via the KOS (which—excepting ontologies—is the rule), the average number of documents per concept is a good estimate for the optimal number of terms in the KOS. Of further interest is the number of designations (synonyms and quasi-synonyms) per concept. The average number of designations (e.g. non-descriptors) of a concept (e.g. of a descriptor) is a good indicator for the use of designations in the KOS.

Analogously to the concepts, the number of different semantic relations used provides an indicator for the structure of a KOS (semantic expressiveness). The total number of relations in the KOS is of particular interest. Regarded as a network, the KOS's concepts represent the nodes while their relations represent the lines. The size of relations is the total number of all lines in the KOS (without the connections to the designations, since these form their own indicator). A useful derived parameter is the average number of semantic relations per concept, i.e. the terms' mean degree. The indicators for size of concepts and size of relations can be summarized as the "granularity of a KOS."

Information concerning the number of hierarchy levels as well as the distribution of terms throughout these individual levels is of particular interest. Also important are data concerning the number of top terms (and thus the different facets) and bottom terms

(concepts on the lowest hierarchy level), each in relation to the total number of all terms in the KOS. The relation of the number of top terms to the number of all terms is called the "fan-out factor," while the analogous relation to the bottom terms can be referred to as "groundedness." "Tangledness" in turn measures the degree of polyhierarchy in the KOS. It refers to the average number of hyperonyms for every concept. By counting the number of hyponyms for all concepts that have hyponyms (minus one), we glean a value for each concept's average number of siblings.

Soergel (2001) proposes measuring the degree of a term's precombination. A KOS's degree of precombination is the average number of partial terms per concept. The degree of precombination for *Garden* is 1, for *Garden Party* it is 2, for *Garden Party Dinner* 3, etc. For English-language KOSs, it is (more or less) easy to count the words forming a term, for other languages, e.g., German with many compounds (*Garten*: 1; *Gartenfest*: 2; *Gartenfestessen*: 3), we have to apply compound decomposition in the first place and then to count for every KOS entry its number of partial terms. Table 2 presents an overview of the multitude of basic KOS structure measures.

**Table 2:** Basic KOS Structure Measures. *Source:* Stock & Stock, 2013, p. 815 (modified).

| Dimension | Informetric Measure | Calculation |
|---|---|---|
| Granularity | Size of concepts | Number of concepts (nodes in the network) |
| | Size of relations | Number of relations between concepts (lines in the network) |
| | Semantic expressiveness | Number of different semantic relations |
| | Documents per concept | Average number of documents per concept (for a given information service) |
| | Use of denotations | Average number of denotations per concept |
| Hierarchy | Depth of hierarchy | Number of levels |
| | Hierarchical distribution of concepts | Number of concepts on the different levels |
| | Fan-out factor | Quotient of the number of top terms and the number of all concepts |
| | Groundedness factor | Quotient of the number of bottom terms and the number of all concepts |
| | Tangledness factor | Average number of hyperonyms per concept |
| | Siblinghood factor | Average number of co-hyponyms per concept |
| Precombination | Degree of precombination | Average number of partial concepts per concept |

*Completeness Indicator*

Completeness refers to the degree of terminological coverage of a knowledge domain. If the knowledge domain is not very small and easily grasped, this value will be very difficult to determine. Yu, Thorn and Tam (2009, p. 775) define completeness via the question: Does the KOS "have concepts missing with regards to the relevant frames of reference?" Portaluppi (2007) demonstrates that the completeness of thematic areas of the KOS can be estimated via samples from indexed documents. In the case study, articles on chronobiology were researched in *Medline*. The original documents were acquired, and the allocated MeSH concepts ("Medical Subject Headings," which is a thesaurus for medical terminology) were analyzed in the surrogates. Portaluppi (2007, p. 1213) reports, "By reading each article, it was ... possible to identify common chronobiologic concepts not yet associated with specific MeSH headings." The missing concepts thus identified might be present in MeSH and may have been erroneously overlooked by the indexer (in which case it would be an indexing error), or they are simply not featured in the KOS. In the case study, some common chronobiological concepts are "not to be associated with any specific MeSH heading" (Portaluppi, 2007, p. 1213), so that MeSH must be deemed incomplete from a chronobiological perspective.

If one counts the concepts in the KOS's thematic subset and determines the number of terms that are missing from a thematic point of view, the quotient of the number of missing terms and the total number of terms (i.e., those featured in the KOS plus those missing) results in an estimated value of completeness or recall (in the sense of Brewster, Alani, Dasmahapatra, & Wilks, 2004) with regard to the corresponding knowledge subdomain.

*Semantic Indicators*

The consistency of a KOS relates to five aspects:

- Semantic inconsistency;
- Circularity error;
- Skipping hierarchical levels;
- Redundancy; and
- the "tennis problem."

Inconsistencies are particularly likely to arise when several KOSs (that are consistent in themselves) are unified into a large KOS. In the case of semantic inconsistency, terms have

been wrongly arranged in the semantic network of all concepts. Consider the following descriptor entry:

Fishes

BT: Marine animals

NT: Salt-water fishes

NT: Freshwater fishes.

BT (broader term) and NT (narrower term) span the semantic relation of hyponymy in the example. In this hierarchical relation, the hyponyms inherit all characteristics of their hyperonyms. The term *Marine animals*, for instance, contains the characteristic "lives in the ocean." This characteristic is passed on to the hyponym *Fishes* and onward to its hyponyms *Salt-water fishes* and *Freshwater fishes*. The semantic inconsistency arises in the case of *Freshwater fishes*, as these do not live in the ocean.

Circularity errors occur in the hierarchical relation when one concept appears more than once in a concept ladder (Gómez-Pérez, 2004a): "Circularity errors ... occur when a class is defined as a specialization or generalization of itself" (p. 261). Suppose that two KOSs are merged. Let KOS 1 contain the following set of concepts:

Persons

NT: Travelers

whereas KOS 2 formulates

Travelers

NT: Persons

When both KOSs are merged, the result is a logical circle (example taken from Cross & Pal, 2008).

Skipping errors are the result of hierarchy levels being left out. This error is well described by Aristotle in his *Topics* (2005, Book 6, Ch. 5, p. 479-480). Here, too, we can provide an example:

Capra

NT: Wild goat

NT: Domestic Goat

Wild goat

NT: Domestic goat.

In the biological hierarchy, *Capra* is the broader term for *Wild goat* (Capra aegagrus). *Wild goat*, in turn, is the broader term for *Domestic goat* (Capra hircus). By establishing a direct relation between *Capra* and *Domestic goat*, our KOS skips a hierarchy level. The cause of the skipping error is the erroneous subsumption of NT *Domestic goat* within the concept *Capra*.

A KOS is redundant when a concept appears more than one time in the KOS. Such an error can occur when the concept is integrated in several contexts. In a thesaurus, *Cherry* may be hyponym of *Fruit tree* and hyperonym of *Sour cherry* and *Sweet cherry*. In another facet of the same thesaurus, *Cherry* is a narrower term of *Fruit brandy* and a broader term of *Cherry brandy*. In this example, the second variant is erroneous. *Cherry* has to be removed from the *Brandy*-facet. Instead of this descriptor, an association link between *Cherry* and *Cherry brandy* should be established.

For ontology evaluation, Hartmann et al. (2005, p.17) mention the so called "tennis problem." This is a phenomenon "where related words could occur in two completely different parts of the ontology with no apparent link between them, e.g., 'ball boy' could occur as a descendant of 'male child' and 'tennis ball' as a descendant of 'game equipment,' despite an obvious semantic relation." Indeed, if a KOS only consists of hierarchy, the tennis problem gives ontology engineers a headache. However, every KOS that allows for the use of the association relation is able to relate both concepts:

<div align="center">Ball boy <i>SEE ALSO</i> Tennis ball</div>

(and vice versa). The task for the evaluator is to locate concepts in the KOS with close semantic relations which are not linked via short paths.

*Overlap with Other KOSs*

An approach to study the similarity between KOSs is to count common words and concepts in two KOSs. On the word level, Maedche and Staab (2002, p. 254) use the Levenshtein distance (i.e., the number of edit steps between two strings). Words with low numbers of editing steps are considered similar. If KOS 1 has the entry "TopHotel" and KOS 2 "Top_Hotel," the Levenshtein distance is 1 (one insertion operation) and the words are therefore similar. But this method is prone to failure. The Levenshtein distance between "Power" and "Tower" is 1 as well despite their dissimilarity. On the concept level, the comparison is even more challenging. Obrst et al. (2007) describe this problem:

> To say that two concepts have similar semantics ... means roughly that they occupy similar places in their lattices. A problem with the above is, however, clear: ontology alignment is defined in terms of correspondence (equivalence, sameness, similarity) of concepts. But how, precisely, do we gain access to concepts in order to determine whether they stand in a relation correspondence? (p. 146)

Obrst et al. (2007) found that the majority of studies are based on the vocabulary (i.e., the words—with the above-mentioned problems) or on the structure of the KOS (e.g., similar broader terms and similar narrower terms). Counting common words and common concepts is a good idea on a theoretical level, but when it comes to practical application, a problem arises.

Fortunately, there is an alternative method. In the case of polyrepresentation (Ingwersen & Järvelin, 2005, p. 346), different methods of knowledge representation as well as different KOSs are used to index the same documents. Haustein and Peters (2012) compare the tags (i.e., in the sense of folksonomies, the readers' perspective), subject headings of *Inspec* (the indexers' perspective), *KeyWords Plus* (as a method of automatic indexing) as well as author keywords and the words from title and abstract (the authors' perspective) of over 700 journal articles. The authors are particularly interested in the overlap between folksonomy-based tags and other methods of knowledge representation. Of course, one can also compare several KOSs with each other, as long as they have been used to index the same documents. The value $g$ represents the number of identical concepts of different KOSs per document, $a$ is the number of unique concepts from KOS 1 per document, and $b$ the number of unique concepts from KOS 2 per document. The similarity of KOS 1 and KOS 2 can be calculated by the Cosine.

The Haustein-Peters method can be used to comparatively evaluate different KOSs in the context of polyrepresentation. When the similarity measurements between two KOSs are relatively low, this points to vocabularies that complement each other—which is of great value to the users, as it provides additional access points to the document. If similarities are high, on the other hand, one of the two KOSs will probably become redundant in that practice.

*Use*

We have learned from Noy (2004) that it is essential for KOS evaluation to consider the KOS users' view. Accordingly, the KOS evaluation has to be embedded into a broader frame, which includes the service, the user, his/her acceptance, the environment and time (Schumann & Stock, 2014). Aspects of user-driven methods include indicators of perceived service quality (captured, e.g., by the SERVQUAL method), perceived system quality with the sub-dimensions of perceived ease of use, usefulness, trust, fun and further factors (applying the Technology Acceptance Model), and usability.

For evaluating the perceived service quality we propose to use SERVQUAL (Parasuraman, Zeithaml, & Berry, 1988). SERVQUAL works with two sets of statements: those that are used to measure expectations about a service category in general (EX) and those that measure perceptions (PE) about the category of a particular service. Each statement is accompanied by a seven-point scale ranging from "strongly disagree" (1) to "strongly agree" (7). For the expectation value, one might note that "in a KOS in economics it is useful to have the relation *has_subsidiary_company* when formulating queries," and then ask the test subject to express this numerically on the given scale. The corresponding statement for registering the perception value would then be: "In the KOS X, the relation *has_subsidiary_company* is useful when formulating queries." Here, too, the subject specifies a numerical value. For each item, a difference score Q = PE – EX is defined. If, for instance, a test subject specifies a value of 1 for perception after having noted a 4 for expectation, the Q value for system X with regard to the attribute in question will be 1 – 4 = -3.

When evaluating perceived KOS quality, questionnaires are used. The test subjects must be familiar with the system in order to make correct assessments. For each subdimension, a set of statements is formulated that the user must estimate on a 7-point scale (from „extremely likely" to „extremely unlikely"). Davis (1989, p. 340), for instance, posited: "using system X in my job would enable me to accomplish tasks more quickly" (to measure perceived usefulness), and "my interaction with system X would be clear and understandable" (for the aspect of perceived ease of use).

*Usable* KOSs are those that do not frustrate the users. A common procedure in usability tests according to Nielson (1993) is task-based testing. Here, an examiner defines representative tasks that can be performed using the KOS and which are typical for such KOSs. Such a task for evaluating the usability of a KOS in economics might be as follows: "Look for concepts to prepare a query on the Fifth Kondratiev cycle!" Test subjects should be "a representative sample of end users" (Rubin & Chisnell, 2008, p. 25). The test subjects are presented with the tasks and are observed by the examiner while they perform them. It is useful to have test subjects speak their thoughts when performing the tasks ("thinking aloud"). In addition to the task-based tests, it is useful for the examiner to interview the subjects on the KOS (e.g., on their overall impression of the KOS, on completeness, and semantic consistency). In Table 3, all mentioned KOS quality indicators are listed.

**Table 3:** KOS Quality Indicators. *Source:* Stock & Stock, 2013, p. 815 (modified).

| Dimension | Informetric indicator | Calculation / method |
|---|---|---|
| Completeness | Completeness of knowledge subdomain | Quotient of the number of missing concepts and the number of all concepts (in the KOS and the missing ones) regarding the subdomain |
| Semantics | Semantic inconsistency | Number of semantic inconsistency errors |
|  | Circularity | Number of circularity errors |
|  | Skipping hierarchical levels | Number of skipping errors |
| Redundancy | Redundancy | Number of redundancy errors |
|  | Tennis problem | Number of missing links between associated concepts |
| Multiple KOSs | Degree of polyrepresentation | Overlap |
| Use | Perceived KOS quality | SERVQUAL questionnaires |
|  | KOS acceptance | Technology acceptance surveys |
|  | Usability | Task-based tests |

**Conclusion**

Our parameters in the group of "Basic Structure" are simple measures, which can be made automatically available by the system. Indeed, it is a quality aspect of every KOS construction and maintenance software to provide such basic structure data (Vogel, 2002). Completeness, semantic consistency, the overlap with other KOSs and user-based data are quality indicators, which "will remain a task for a human level intelligence" (Vrandečić, 2009, p. 308).

Next steps in KOSs evaluation research should include the analysis of the different evaluation methods. Gómez-Pérez (2004b, p. 74) mentions research questions such as "How robust are ontology evaluation methods?" or "How do ontology development platforms perform content evaluation?"

In this chapter, we tried to focus information scientists' attention to a widely neglected aspect of informetrics: the informetric description and evaluation of KOSs. As a basis for further discussion, we described the state of the art of KOS evaluation and introduced suggestions for measures as well as indicators of the quality of KOSs. We hope that we were able to expand the theory of informetrics by introducing evaluation methods of KOSs.

# Cited References